\documentclass[journal]{IEEEtran}
\usepackage{url}
\usepackage[utf8]{inputenc}
\usepackage{xcolor}
\usepackage{amsmath}
\usepackage{amssymb}

\usepackage[acronyms,nonumberlist,nopostdot,nomain,nogroupskip]{glossaries}
\usepackage{tablefootnote}
\usepackage{booktabs}
\usepackage{tabularx}
\usepackage{epsfig}
\usepackage[outdir=img/]{epstopdf}
\usepackage{tikz}
\usepackage{pgfplots}
\pgfplotsset{compat=newest} 
\pgfplotsset{plot coordinates/math parser=false} 
\newlength\fheight
\newlength\fwidth
\usetikzlibrary{plotmarks,shapes,patterns,decorations.pathreplacing,backgrounds,calc,arrows,arrows.meta,spy,matrix}
\usepgfplotslibrary{patchplots,groupplots}
\usepackage{tikzscale}
\usepackage{hyperref}
\usepackage{physics}

\usepackage{multirow}
\usepackage{tkz-kiviat}

\usepackage[font=scriptsize]{subcaption}
\usepackage[font=footnotesize]{caption}

\usepackage{mathtools}

\usepackage{dblfloatfix}    
\usepackage{colortbl}

\newacronym{3gpp}{3GPP}{3rd Generation Partnership Project}
\newacronym{adc}{ADC}{Analog to Digital Converter}
\newacronym{afbw}{AFBW}{Average Fading Bandwidth}
\newacronym{5g}{5G}{5th generation}
\newacronym{aimd}{AIMD}{Additive Increase Multiplicative Decrease}
\newacronym{am}{AM}{Acknowledged Mode}
\newacronym{amc}{AMC}{Adaptive Modulation and Coding}
\newacronym{aqm}{AQM}{Active Queue Management}
\newacronym{awgn}{AGWN}{Additive White Gaussian Noise}
\newacronym{balia}{BALIA}{Balanced Link Adaptation}
\newacronym{bdp}{BDP}{Bandwidth-Delay Product}
\newacronym{bf}{BF}{Beamforming}
\newacronym{cc}{CC}{Congestion Control}
\newacronym{cdf}{CDF}{Cumulative Distribution Function}
\newacronym{cn}{CN}{Core Network}
\newacronym{cqi}{CQI}{Channel Quality Information}
\newacronym{cir}{CIR}{Channel Impulse Response}
\newacronym{cp}{CP}{Control Plane}
\newacronym{csirs}{CSI-RS}{Channel State Information - Reference Signal}
\newacronym{dc}{DC}{Dual Connectivity}
\newacronym{dce}{DCE}{Direct Code Execution}
\newacronym{dci}{DCI}{Downlink Control Information}
\newacronym{dl}{DL}{Downlink}
\newacronym{dmr}{DMR}{Deadline Miss Ratio}
\newacronym{dmrs}{DMRS}{DeModulation Reference Signal}
\newacronym{e2e}{E2E}{End-to-End}
\newacronym{ecn}{ECN}{Explicit Congestion Notification}
\newacronym{edf}{EDF}{Earliest Deadline First}
\newacronym{enb}{eNB}{evolved Node Base}
\newacronym{epc}{EPC}{Evolved Packet Core}
\newacronym{es}{ES}{Edge Server}
\newacronym{fdma}{FDMA}{Frequency Division Multiple Access}
\newacronym{fdd}{FDD}{Frequency Division Duplexing}
\newacronym[firstplural=Radio Access Technologies (RATs)]{rat}{RAT}{Radio Access Technology}
\newacronym{fs}{FS}{Fast Switching}
\newacronym{ftp}{FTP}{File Transfer Protocol}
\newacronym{gnb}{gNB}{Next Generation Node Base}
\newacronym{harq}{HARQ}{Hybrid Automatic Repeat reQuest}
\newacronym{hetnet}{HetNet}{Heterogeneous Network}
\newacronym{hh}{HH}{Hard Handover}
\newacronym{hol}{HOL}{Head-of-Line}
\newacronym{ia}{IA}{Initial Access}
\newacronym{imt}{IMT}{International Mobile Telecommunication}
\newacronym{iot}{IoT}{Internet of Things}
\newacronym{lcr}{LCR}{Level Crossing Rate}
\newacronym{lcf}{LCF}{Level Crossing Frequency}
\newacronym{los}{LoS}{Line-of-Sight}
\newacronym{lte}{LTE}{Long Term Evolution}
\newacronym{m2m}{M2M}{Machine to Machine}
\newacronym{mac}{MAC}{Medium Access Control}
\newacronym{mc}{MC}{Multi-Connectivity}
\newacronym{mcs}{MCS}{Modulation and Coding Scheme}
\newacronym{mec}{MEC}{Mobile Edge Cloud}
\newacronym{mi}{MI}{Mutual Information}
\newacronym{mimo}{MIMO}{Multiple Input, Multiple Output}
\newacronym{mmwave}{mmWave}{millimeter wave}
\newacronym{moi}{MoI}{Method of Images}
\newacronym{mptcp}{MPTCP}{Multipath TCP}
\newacronym{mr}{MR}{Maximum Rate}
\newacronym{mss}{MSS}{Maximum Segment Size}
\newacronym{mtd}{MTD}{Machine-Type Device}
\newacronym{mtu}{MTU}{Maximum Transmission Unit}
\newacronym{nfv}{NFV}{Network Function Virtualization}
\newacronym{nlos}{NLoS}{Non-Line-of-Sight}
\newacronym{nr}{NR}{New Radio}
\newacronym{o2i}{O2I}{Outdoor-to-Indoor}
\newacronym{ofdm}{OFDM}{Orthogonal Frequency Division Multiplexing}
\newacronym{pdcch}{PDCCH}{Physical Downlonk Control Channel}
\newacronym{pdcp}{PDCP}{Packet Data Convergence Protocol}
\newacronym{pdsch}{PDSCH}{Physical Downlink Shared Channel}
\newacronym{pdu}{PDU}{Packet Data Unit}
\newacronym{pf}{PF}{Proportional Fair}
\newacronym{pgw}{PGW}{Packet Gateway}
\newacronym{phy}{PHY}{Physical}
\newacronym{pbch}{PBCH}{Physical Broadcast Channel}
\newacronym[plural=\gls{mme}s,firstplural=Mobility Management Entities (MMEs)]{mme}{MME}{Mobility Management Entity}
\newacronym{prb}{PRB}{Physical Resource Block}
\newacronym{pss}{PSS}{Primary Synchronization Signal}
\newacronym{pucch}{PUCCH}{Physical Uplink Control Channel}
\newacronym{pusch}{PUSCH}{Physical Uplink Shared Channel}
\newacronym{rach}{RACH}{Random Access Channel}
\newacronym{ran}{RAN}{Radio Access Network}
\newacronym{red}{RED}{Random Early Detection}
\newacronym{rf}{RF}{Radio Frequency}
\newacronym{rlc}{RLC}{Radio Link Control}
\newacronym{rlf}{RLF}{Radio Link Failure}
\newacronym{rrc}{RRC}{Radio Resource Control}
\newacronym{rrm}{RRM}{Radio Resource Management}
\newacronym{rr}{RR}{Round Robin}
\newacronym{rs}{RS}{Remote Server}
\newacronym{rsrp}{RSRP}{Reference Signal Received Power}
\newacronym{rss}{RSS}{Received Signal Strength}
\newacronym{rtt}{RTT}{Round Trip Time}
\newacronym{rw}{RW}{Receive Window}
\newacronym{rx}{RX}{Receiver}
\newacronym{sa}{SA}{standalone}
\newacronym{sack}{SACK}{Selective Acknowledgment}
\newacronym{sap}{SAP}{Service Access Point}
\newacronym{sch}{SCH}{Secondary Cell Handover}
\newacronym{scoot}{SCOOT}{Split Cycle Offset Optimization Technique}
\newacronym{sdma}{SDMA}{Spatial Division Multiple Access}
\newacronym{sinr}{SINR}{Signal-to-Interference-plus-Noise Ratio}
\newacronym{sir}{SIR}{Signal-to-Interference Ratio}
\newacronym{sm}{SM}{Saturation Mode}
\newacronym{snr}{SNR}{Signal-to-Noise Ratio}
\newacronym{son}{SON}{Self-Organizing Network}
\newacronym{ss}{SS}{Synchronization Signal}
\newacronym{srs}{SRS}{Sounding Reference Signal}
\newacronym{sss}{SSS}{Secondary Synchronization Signal}
\newacronym{tb}{TB}{Transport Block}
\newacronym{tcp}{TCP}{Transmission Control Protocol}
\newacronym{tdd}{TDD}{Time Division Duplexing}
\newacronym{tdma}{TDMA}{Time Division Multiple Access}
\newacronym{tfl}{TfL}{Transport for London}
\newacronym{tm}{TM}{Transparent Mode}
\newacronym{trp}{TRP}{Transmitter Receiver Pair}
\newacronym{tti}{TTI}{Transmission Time Interval}
\newacronym{ttt}{TTT}{Time-to-Trigger}
\newacronym{tx}{TX}{Transmitter}
\newacronym{ue}{UE}{User Equipment}
\newacronym{ul}{UL}{Uplink}
\newacronym{uml}{UML}{Unified Modeling Language}
\newacronym{um}{UM}{Unacknowledged Mode}
\newacronym{uma}{UMa}{Urban Macro}
\newacronym{utc}{UTC}{Urban Traffic Control}
\newacronym{vm}{VM}{Virtual Machine}
\newacronym{rsrq}{RSRQ}{Reference Signal Received Quality}
\newacronym{rssi}{RSSI}{Received Signal Strength Indicator}
\newacronym{crs}{CRS}{Cell Reference Signal}
\newacronym{nsa}{NSA}{Non Stand Alone}
\newacronym{mrdc}{MR-DC}{Multi \gls{rat} \gls{dc}}
\newacronym{endc}{EN-DC}{E-UTRAN-\gls{nr} \gls{dc}}
\newacronym{5gc}{5GC}{5G Core}
\newacronym{si}{SI}{Study Item}
\newacronym{iab}{IAB}{Integrated Access and Backhaul}
\newacronym{wf}{WF}{Wired-first}
\newacronym{hqf}{HQF}{Highest-quality-first}
\newacronym{pa}{PA}{Position-aware}
\newacronym{mlr}{MLR}{Maximum-local-rate}
\newacronym{wbf}{WBF}{Wired Bias Function}
\newacronym{mib}{MIB}{Master Information Block}
\newacronym{sib}{SIB}{Secondary Information Block}
\newacronym{kpi}{KPI}{Key Performance Indicator}
\newacronym{ppp}{PPP}{Poisson Point Process}
\newacronym{mpc}{MPC}{Multi Path Component}
\newacronym{rt}{RT}{Ray Tracer}
\newacronym{aoa}{AoA}{Angle of Arrival}
\newacronym{aod}{AoD}{Angle of Departure}
\newacronym{scm}{SCM}{Spatial Channel Model}
\newacronym{inr}{INR}{Interference to Noise Ratio}
\newacronym{qd}{QD}{Quasi Deterministic}
\newacronym{wlan}{WLAN}{Wireless Local Area Network}
\newacronym{cad}{CAD}{Computer-aided Design}
\newacronym{ap}{AP}{Access Point}
\newacronym{sta}{STA}{Station}
\newacronym{nrmse}{NRMSE}{Normalized Root Mean Square Error}
\tikzstyle{startstop} = [rectangle, rounded corners, minimum width=2cm, minimum height=0.5cm,text centered, draw=black]
\tikzstyle{io} = [trapezium, trapezium left angle=70, trapezium right angle=110, minimum width=3cm, minimum height=1cm, text centered, draw=black]
\tikzstyle{process} = [rectangle, minimum width=2cm, minimum height=0.5cm, text centered, draw=black, alignb=center]
\tikzstyle{decision} = [ellipse, minimum width=2cm, minimum height=1cm, text centered, draw=black]
\tikzstyle{arrow} = [thick,<->,>=stealth]
\tikzstyle{line} = [thick,>=stealth]
\tikzstyle{darrow} = [thick,<->,>=stealth,dashed]
\tikzstyle{sarrow} = [thick,->,>=stealth]
\tikzstyle{larrow} = [line width=0.1mm,dashdotted,->,>=stealth]

\makeatletter
\def\grd@save@target#1{%
  \def\grd@target{#1}}
\def\grd@save@start#1{%
  \def\grd@start{#1}}
\tikzset{
  grid with coordinates/.style={
    to path={%
      \pgfextra{%
        \edef\grd@@target{(\tikztotarget)}%
        \tikz@scan@one@point\grd@save@target\grd@@target\relax
        \edef\grd@@start{(\tikztostart)}%
        \tikz@scan@one@point\grd@save@start\grd@@start\relax
        \draw[minor help lines] (\tikztostart) grid (\tikztotarget);
        \draw[major help lines] (\tikztostart) grid (\tikztotarget);
        \grd@start
        \pgfmathsetmacro{\grd@xa}{\the\pgf@x/1cm}
        \pgfmathsetmacro{\grd@ya}{\the\pgf@y/1cm}
        \grd@target
        \pgfmathsetmacro{\grd@xb}{\the\pgf@x/1cm}
        \pgfmathsetmacro{\grd@yb}{\the\pgf@y/1cm}
        \pgfmathsetmacro{\grd@xc}{\grd@xa + \pgfkeysvalueof{/tikz/grid with coordinates/major step x}}
        \pgfmathsetmacro{\grd@yc}{\grd@ya + \pgfkeysvalueof{/tikz/grid with coordinates/major step y}}
        \foreach \x in {\grd@xa,\grd@xc,...,\grd@xb}
        \node[anchor=north] at (\x,\grd@ya) {\pgfmathprintnumber{\x}};
        \foreach \y in {\grd@ya,\grd@yc,...,\grd@yb}
        \node[anchor=east] at (\grd@xa,\y) {\pgfmathprintnumber{\y}};
      }
    }
  },
  minor help lines/.style={
    help lines,
    gray,
    line cap =round,
    xstep=\pgfkeysvalueof{/tikz/grid with coordinates/minor step x},
    ystep=\pgfkeysvalueof{/tikz/grid with coordinates/minor step y}
  },
  major help lines/.style={
    help lines,
    line cap =round,
    line width=\pgfkeysvalueof{/tikz/grid with coordinates/major line width},
    xstep=\pgfkeysvalueof{/tikz/grid with coordinates/major step x},
    ystep=\pgfkeysvalueof{/tikz/grid with coordinates/major step y}
  },
  grid with coordinates/.cd,
  minor step x/.initial=.5,
  minor step y/.initial=.2,
  major step x/.initial=1,
  major step y/.initial=1,
  major line width/.initial=1pt,
}
\makeatother

\IEEEoverridecommandlockouts
\newcommand\copyrighttext{%
  \footnotesize \textcopyright 2020 IEEE. Personal use of this material is permitted.
  Permission from IEEE must be obtained for all other uses, in any current or future media, including reprinting/republishing this material for advertising or promotional purposes, creating new collective works, for resale or redistribution to servers or lists, or reuse of any copyrighted component of this work in other works.}
\newcommand\copyrightnotice{%
\begin{tikzpicture}[remember picture,overlay]
\node[anchor=south,yshift=5pt] at (current page.south) {\fbox{\parbox{\dimexpr\textwidth-\fboxsep-\fboxrule\relax}{\copyrighttext}}};
\end{tikzpicture}%
}

\usepackage[capitalize]{cleveref}

\makeglossaries

\begin{document}

\title{Simplified Ray Tracing for the Millimeter Wave Channel: A Performance Evaluation }

\author{\IEEEauthorblockN{Mattia Lecci\IEEEauthorrefmark{1},
                          Paolo Testolina\IEEEauthorrefmark{1},
                          Marco Giordani\IEEEauthorrefmark{1},
                          Michele Polese\IEEEauthorrefmark{1},\\
                          Tanguy Ropitault\IEEEauthorrefmark{2},
                          Camillo Gentile\IEEEauthorrefmark{2},
                          Neeraj Varshney\IEEEauthorrefmark{2},
                          Anuraag Bodi\IEEEauthorrefmark{2},
                          Michele Zorzi\IEEEauthorrefmark{1}}\\
        \IEEEauthorblockA{\vspace{0.2cm}    \small \IEEEauthorrefmark{1}Department of Information Engineering, University of Padova, Italy, email: \texttt{\{name.surname\}@dei.unipd.it}\\
        \small \IEEEauthorrefmark{2}National Institute of Standards and Technology (NIST), Gaithersburg, MD, 20899 USA, email: \texttt{\{name.surname\}@nist.gov}}
        \vspace{-5ex}%
        \thanks{This work was partially supported by NIST under Award No. 70NANB18H273.
        Mattia Lecci and Paolo Testolina's activities were supported by \textit{Fondazione CaRiPaRo} under the grants ``Dottorati di Ricerca'' 2018 and 2019, respectively.}
        \thanks{The identification of any commercial product or trade name does not imply endorsement or recommendation by the National Institute of Standards and Technology, nor is it intended to imply that the materials or equipment identified are necessarily the best available for the purpose.}
}

\maketitle
\copyrightnotice

\glsunset{nr}

\begin{abstract}
Millimeter-wave (mmWave) communication is one of the cornerstone innovations of fifth-generation (5G) wireless networks, thanks to the massive bandwidth available in these frequency bands. To correctly assess the performance of such systems, however, it is essential to have reliable channel models, based on a deep understanding of the propagation characteristics of the mmWave signal. In this respect, ray tracers can provide high accuracy, at the expense of a significant computational complexity, which limits the scalability of simulations. To address this issue, in this paper we present possible simplifications that can reduce the complexity of ray tracing in the mmWave environment, without significantly affecting the accuracy of the model. We evaluate the effect of such simplifications on link-level metrics, testing different configuration parameters and propagation scenarios.
\vspace{-3ex}
\end{abstract}

\begin{tikzpicture}[remember picture,overlay]
\node[anchor=north,yshift=-10pt] at (current page.north) {\parbox{\dimexpr\textwidth-\fboxsep-\fboxrule\relax}{\centering \footnotesize This paper has been accepted for presentation at ITA 2020. \textcopyright 2020 IEEE.\\
  Please cite it as: M. Lecci, P. Testolina, M. Giordani, M. Polese, T. Ropitault, C. Gentile, N. Varshney, A. Bodi, M. Zorzi, ``Simplified Ray Tracing for the Millimeter Wave Channel: A Performance Evaluation,'' Information Theory and Applications Workshop (ITA), San Diego, US, 2020}};
\end{tikzpicture}%

\section{Introduction}

The next generation of Cellular and \glspl{wlan} will be the first to exploit \gls{mmwave} frequencies to provide connectivity in the access network, i.e., in the links between base stations and mobile users. In particular, \gls{3gpp} NR has been designed to support a carrier frequency up to 52.6 GHz in Release 15~\cite{38300}, and future Releases will consider extensions to 71 GHz and to the sub-THz band~\cite{qualcomm201971}. Similarly, IEEE 802.11ad and 802.11ay exploit the unlicensed bands at 60 GHz~\cite{802.11ad-standard}. The \gls{mmwave} frequencies, indeed, feature large chunks of untapped bandwidth that can increase the data rate provided to the end-users, making it possible to target the \gls{5g} requirements of ultra-high peak throughput (20 Gbps) and average user experienced rate (50-100 Mbps)~\cite{itu-r-2083}. Moreover, the small wavelength enables the design of antenna arrays with tens of elements in a small form factor, which could fit even smartphones or VR headsets.

The propagation characteristics of the \gls{rf} signals in these frequency bands, however, complicate the design of reliable communication systems~\cite{rangan2017potentials}. First, the high propagation loss, which is proportional to the square of the carrier frequency, limits the coverage region of the mmWave base stations. This can be compensated by using beamforming with large antenna arrays, which could concentrate the power in narrow, directional beams and increase the link budget.
Additionally, \gls{mmwave} signals can  be easily blocked by obstacles (e.g., vehicles, buildings, human bodies), preventing direct \gls{los} communications. Furthermore, at \gls{mmwave} frequencies, the increased diffraction loss results in deep shadow regions, thus further degrading the propagation performance~\cite{dengMmwaveDiffraction}. 
By considering the combination of these phenomena, the \gls{mmwave} channel appears extremely volatile to mobile users, whose quality of experience may be poor unless a proper network design is adopted.

\subsection{Channel Modeling for mmWaves}

As experimental platforms and testbeds at \glspl{mmwave} are still at an early development stage~\cite{polese2019millimetera,saha2019x60}, analysis and simulation play a fundamental role in the performance evaluation of novel solutions for \gls{mmwave} networks. Given the aforementioned behavior of the channel at such high frequencies, and the interplay with network deployment choices and beamforming design, the accuracy of analysis and simulation depends on that of the channel model even more than at conventional sub-6 GHz frequencies. Therefore, the research community has developed a number of channel modeling tools for \glspl{mmwave}, with a varying degree of complexity and accuracy. Stochastic and analytical models are based on the combination of random variables fitted on traces and measurements, and are widely used for analysis~\cite{bai2015coverage,andrews2017modeling} and system-level simulations~\cite{3gpp.38.901,mezzavilla2018end}. However, the generality of these models and their stochastic nature fit poorly with the need to accurately characterize specific scenarios. Additionally, most stochastic models may not properly characterize the specific features of the \gls{mmwave} channel that may affect the overall system performance, such as the temporally- and spatially-consistent updates of the \gls{los} condition and the evolution of the channel \glspl{mpc}, each representing a distinct planar wavefront propagating between the \gls{tx} and the \gls{rx}.

These modeling challenges are instead addressed by \glspl{rt}, which have been used to precisely characterize the propagation of \gls{rf} signals in specific scenarios~\cite{degliesposti2014rt,larew2013air,maltsev2016channel}. With ray tracing, the channel is modeled in terms of \glspl{mpc} that are generated from a certain location and angle of departure, are reflected (and, in complete models, diffracted and diffused) on the scattering surfaces of the scenarios, and reach the position of the receiver with a certain angle of arrival, delay, and power~\cite{gentile2019qdLectureRoom}. As the generation of \glspl{mpc} is purely based on the geometry of the scenario, the channel is as accurate as the description of the scenario, and the \glspl{mpc} are consistent with the mobility model of the communication endpoints. Additionally, ray tracers can be easily integrated into system-level simulators, by computing the channel matrix $\mathbf{H}$ that combines the different multipath components and the antenna arrays of the network nodes.

With currently available channel modeling tools, however, a higher accuracy translates into a higher computational complexity for the \gls{mpc} generation and the simulations. As we discuss in~\cite{testolina2019scalable}, the complexity is proportional to two elements, i.e., the number of \glspl{mpc} which need to be combined to generate the channel matrix $\mathbf{H}$, and the number of antennas at the two endpoints of the communication link (which represents the number of columns and rows of $\mathbf{H}$). 

\subsection{Contributions}

Based on the above introduction, in this paper we investigate whether it is possible to improve the trade-off between accuracy and complexity in mmWave simulations, by studying simplification techniques for ray tracers that speed up the simulations and the ray tracer itself. Specifically, we consider processing only \glspl{mpc} whose received power is above a certain threshold, which is relative to the strongest \gls{mpc}, and limiting the maximum number of reflections for each \gls{mpc}.
Our results show that it is possible to decrease the complexity of the simulations with a minimal reduction in accuracy, with respect to the baseline ray tracer implementation (i.e., without simplifications).

These promising results are a first step towards understanding and isolating which are the most fundamental channel modeling components at \gls{mmwave} frequencies, and could stimulate further investigations into whether it is possible to develop simplified channel models (e.g., to be used also for mathematical analysis) that are more representative of the \gls{mmwave} propagation than the widely used Nakagami-m or Rayleigh fading models~\cite{polese2018impact}.

The rest of the paper is organized as follows. In \cref{sec:rt} we provide details on the ray tracer we consider as the baseline. We then introduce possible simplifications in \cref{sec:simplify}, while performance results are discussed in \cref{sec:results}. Finally, we conclude the paper in \cref{sec:conclusions}.

\section{The Millimeter Wave Ray Tracer}
\label{sec:rt}
To simulate a realistic channel, an open-source MATLAB ray tracer was used\footnote{Ray tracer implementation: \url{https://github.com/wigig-tools/qd-realization}}.
The ray tracer was built with \gls{mmwave} propagation in mind and for this reason, given the deep shadow effect that diffraction yields at such high frequencies~\cite{dengMmwaveDiffraction}, only specular reflections are considered.
In this work, diffuse reflections are ignored, but their importance is undoubted and will thus be part of our future analysis.
Currently, the ray tracer accepts \gls{cad} files in the Additive Manufacturing File (AMF) format, with scenarios represented as a collection of objects in the environment, each object associated with material properties and defined geometrically by a set of triangles\footnote{Triangles are typically used in computer graphics to model physical objects.}.

\begin{figure}[t]
  \centering
  \setlength\fwidth{0.8\columnwidth} 
  \setlength\fheight{0.5\columnwidth}
%
%
\begin{tikzpicture}

\begin{axis}[%
width=\fwidth,
height=\fheight,
at={(0\fwidth,0\fheight)},
scale only axis,
xmin=-8,
xmax=13,
xtick={\empty},
ytick={\empty},
ymin=-5,
ymax=13,
axis background/.style={fill=white},
hide axis,
legend style={legend cell align=left, align=left, draw=white!15!black}
]
\addplot [color=black, line width=1.5pt, forget plot]
  table[row sep=crcr]{%
0	0\\
10	0\\
};
\node[anchor=west] at (10,0) {$S_1$};

\addplot [color=black, line width=1.5pt, forget plot]
  table[row sep=crcr]{%
0	0\\
0	10\\
};
\node[anchor=south west] at (0,10) {$S_2$};

\addplot[only marks, mark=*, mark options={}, mark size=1.5000pt, color=red, fill=red] table[row sep=crcr, forget plot]{%
x	y\\
7	2\\
};
\node[anchor=south west] at (7,2) {$\rm {\color{red}RX} = RX^{(0)}$};

\addplot[only marks, mark=*, mark options={}, mark size=1.5000pt, color=red, fill=red] table[row sep=crcr, forget plot]{%
x	y\\
4	9\\
};
\node[anchor=west] at (4,9) {$\rm {\color{red}TX} = P^{(3)}$};

\addplot[only marks, mark=*, mark options={}, mark size=1.5000pt, color=black, fill=black] table[row sep=crcr, forget plot]{%
x	y\\
7	-2\\
};
\node[anchor=west] at (7,-2) {$\rm RX^{(1)}$};

\addplot[only marks, mark=*, mark options={}, mark size=1.5000pt, color=black, fill=black] table[row sep=crcr, forget plot]{%
x	y\\
-7	-2\\
};
\node[anchor=north west] at (-7,-2) {$\rm RX^{(2)}$};

\addplot [color=black, dashed, line width=.5pt, forget plot]
  table[row sep=crcr]{%
-7	-2\\
4	9\\
};

\addplot[only marks, mark=*, mark options={}, mark size=1.5000pt, color=black, fill=black] table[row sep=crcr, forget plot]{%
x	y\\
0	5\\
};
\node[anchor=south east] at (0,5) {$\rm P^{(2)}$};

\addplot [color=black, dashed, line width=.5pt, forget plot]
  table[row sep=crcr]{%
0	5\\
7	-2\\
};

\addplot[only marks, mark=*, mark options={}, mark size=1.5000pt, color=black, fill=black] table[row sep=crcr, forget plot]{%
x	y\\
5	0\\
};
\node[anchor=north east] at (5,0) {$\rm P^{(1)}$};

\addplot [color=black, dashed, line width=.5pt, forget plot]
  table[row sep=crcr]{%
5	0\\
7	2\\
};

\end{axis}
\end{tikzpicture}%
  \caption{Visualization of the Method of Images algorithm for a second-order reflection ($N=2$).}
  \label{fig:rt_explanation}
  \vspace{-3ex}
\end{figure}
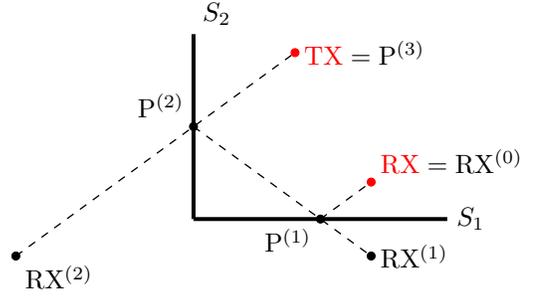

Specular reflections are computed using the \gls{moi}, a basic principle from antenna theory~\cite{raytraceingSurveyYun2015}.
Given two points in 3D space, i.e., the \gls{tx} and the \gls{rx}, and a surface $S$, the \gls{moi} defines the virtual image of the \gls{rx} ($\rm RX^{(1)}$) to be the reflection of the \gls{rx} ($\rm RX^{(0)}$) across the given surface $S$.
By joining the \gls{tx} with the $\rm RX^{(1)}$, it is possible to easily compute the point of specular reflection $\rm P^{(1)}$ as the intersection of the segment with $S$.
It is necessary, though, to check if the reflection point is inside the bounded surface, otherwise, the reflection will be discarded.
Furthermore, it is also necessary to check that every other surface of the scenario does not intersect the two segments at any point, otherwise the whole ray will be considered obstructed and thus discarded.

When multiple reflections are considered, the \gls{moi} applies recursively.
Specifically, given an array of reflecting surfaces $\mathcal{S} = \qty(S_1, \ldots, S_N)$, $\text{RX}^{(n)}$ is computed as the virtual image of $\text{RX}^{(n-1)}$ for surface $S_n, \; n=1, \ldots, N$.
Then, defining $\rm P^{(N+1)} = TX$, the reflection point $\rm P^{(n)}$ is computed as the intersection between the surface $S_n$ and the segment joining $\text{P}^{(n+1)}$ and $\text{RX}^{(n)}$.
Finally, a check is needed on every path segment $\qty(\text{P}^{(n)}, \text{P}^{(n+1)})$ to asses whether it is obstructed by any triangle of the~environment.
Fig.~\ref{fig:rt_explanation} shows a visual example of the \gls{moi} algorithm.

To compute all possible reflections between the \gls{tx} and the \gls{rx}, a \textit{reflection tree} is created, based on the geometric information extracted from the \gls{cad} file.
All the nodes of a reflection tree correspond to triangles of the environment, except the root, that represents the location of the ray source, i.e., the \gls{tx}.
For each node of the reflection tree, its children coincide with all the other triangles of the CAD file.
Thus, the depth of the tree corresponds to the maximum reflection order $\eta_{\rm max}$ (given as an input configuration parameter), i.e., the maximum number of reflections per \gls{mpc} that the RT computes, and each path from the root to a node at depth $d$ corresponds to an ordered array of $d$ reflecting surfaces.
By following all the paths for each tree depth $d$, all possible arrays of triangles are tested and thus every reflected ray is computed.

Accurate profiling of the software shows that the most demanding parts of the \gls{rt} operations are the geometric computations (i.e., computing the position of virtual \glspl{rx}, computing the point of specular reflection, check if the point is inside the bounded surface) and the obstruction checks.
The complexity of the geometric computations is proportional to $\eta_{\rm max}$, while obstruction checks scale both with $\eta_{\rm max}$ (every segment has to be checked) and the environment complexity (any triangle can potentially block the propagation of the ray).

Finally, if the ray's reflections are valid and there are no obstructions, the path gain is computed as
\begin{equation}
  PG = \qty(\frac{\lambda}{4\pi d})^2 - \sum_{n=1}^N RL_n,
\end{equation}
where $\lambda$ is the wavelength (that is a function of the carrier frequency $f_c$), $d$ is the total distance traveled by the ray, and $RL_n$ is the reflection loss of the material associated to the $n$-th reflecting surface~\cite{gentile2019qdLectureRoom}.
Together with the delay, phase, angle of departure, and angle of arrival, the path gain is returned as an output by the ray tracer and written to a file in a specified format, which can be fed as input to other simulators (e.g., link-level or system-level simulators) to compute the channel between the two nodes.

To further simplify the software from the Fresnel equations, which dictate the laws of reflection, no polarization is considered, and rays reflected by a surface experience a $180^\circ$ phase rotation and a reflection loss $RL_n$ between $7$~dB to $25$~dB, depending on the material but irrespective of the angle of incidence.

\section{How to Simplify the Millimeter Wave Channel}
\label{sec:simplify}
The number of multipath components of the channel between two endpoints heavily affects the computational complexity of the \gls{rt}, as outlined in Sec.~\ref{sec:rt}, and of the higher-level simulators~\cite{testolina2019scalable}, e.g., link-level or network-level ones.
For this reason, in this work we propose two different strategies to reduce the total number of \glspl{mpc}, and analyze the effects that this simplification yields on the system behavior.
The two strategies we introduce are based on considerations related to the power of every single \gls{mpc}, based on the idea that weak \glspl{mpc} do not significantly contribute to the overall signal at the receiver.

For the first simplification approach, we reduce the number of reflections $\eta_{\rm max}$ that the \gls{rt} computes. Indeed, when considering an increasing number of reflections, the \gls{mpc} experiences a decreasing path gain $PG$, as the absorption on the reflecting surfaces severely degrades the power of the ray and the length of the ray increases. Therefore, \glspl{mpc} that bounce across multiple scattering surfaces have a low contribution to the overall received power, and can be omitted from the \gls{rt} computations.
Limiting the maximum reflection order corresponds to bounding the depth of the reflection tree, whose size, as mentioned in the previous section, is exponential in the maximum reflections order $\eta_{\rm max}$. Therefore, reducing $\eta_{\rm max}$ reduces the complexity both of the \gls{rt} and of the network simulators that use it to model the channel.

Following similar considerations, the second strategy aims at discarding the weakest \glspl{mpc} based on how low their path gain is compared to the strongest \gls{mpc}, regardless of how many reflections they experience. Using this method, the path gain $PG$ for a ray still needs to be computed, therefore the geometric operations will not be spared. However, the obstruction check may not need to be performed in the case $PG$ is below a certain threshold, and the ray is not accounted for when computing the channel matrix $\mathbf{H}$.
The method we propose is based on a threshold $\gamma_{\rm th}$ which is relative to the $PG_{\rm max}$ of the strongest \gls{mpc} for a given channel realization. Notably, the rays with path gain $PG / PG_{\rm max} < \gamma_{\rm th}$ are discarded. 
The path gain associated with the strongest ray $PG_{\rm max}$ is updated on-line.

Selecting the \glspl{mpc} with a relative, rather than absolute, threshold makes it possible to dynamically adapt the simplification to the actual quality of the channel.
For example, in \gls{nlos} conditions, the strongest \gls{mpc} will be given by a reflected ray.
Therefore, its path gain $PG_{\rm max}$ will be comparable to that of a higher number of \glspl{mpc} (given by other reflections) than in a \gls{los} scenario, where the strongest ray is the direct path, with a much higher power than the reflections.
If the reflections have a $PG$ similar to $PG_{\rm max}$, then the receiver experiences a strong fading.
In this case, a \textit{relative} threshold combines accuracy (in \gls{nlos}, significant \glspl{mpc} are still computed) and reduction in complexity (in \gls{los}, several negligible reflected \glspl{mpc} are not accounted for), more than an absolute threshold, which would apply the same cut in both cases.


We implemented the proposed techniques in the open-source ray tracer described in Sec.~\ref{sec:rt}. Although such methods are beneficial from a computational point of view, they may have some drawbacks, depending on the reliability requirements of the application for which the ray tracer is used. The most evident downside is that the power spatial distribution can be affected in complex and non-foreseeable ways, as we will show in Sec.~\ref{sec:results}.

\section{Performance Results}
\label{sec:results}

In this section, we evaluate through simulations the effects of the ray tracer simplifications we presented in Sec.~\ref{sec:simplify}. In particular, we investigate the impact of (i) the maximum number of reflections $\eta_{\rm max}$ per \gls{mpc}, and (ii)  the received power threshold $\gamma_{\rm th}$ (relative to the strongest path) below which MPCs are discarded by the ray tracer.

\begin{figure*}[tbp]
  \hfill
  \begin{subfigure}[b]{0.3\textwidth}
      \centering
      \setlength\fwidth{1.0\columnwidth} 
      \setlength\fheight{0.5\columnwidth}
      \input{img/Indoor1.tex}
      \caption{\textit{Indoor1}}
      \label{fig:indoor1}
  \end{subfigure}
  \hfill
  \begin{subfigure}[b]{0.3\textwidth}
      \centering
      \setlength\fwidth{1.0\columnwidth} 
      \setlength\fheight{0.5\columnwidth}
      \input{img/L-room.tex}
      \caption{\textit{L-room}}
      \label{fig:l-room}
  \end{subfigure}
  \hfill
  \begin{subfigure}[b]{0.3\textwidth}
        \centering
        \includegraphics[width=\columnwidth]{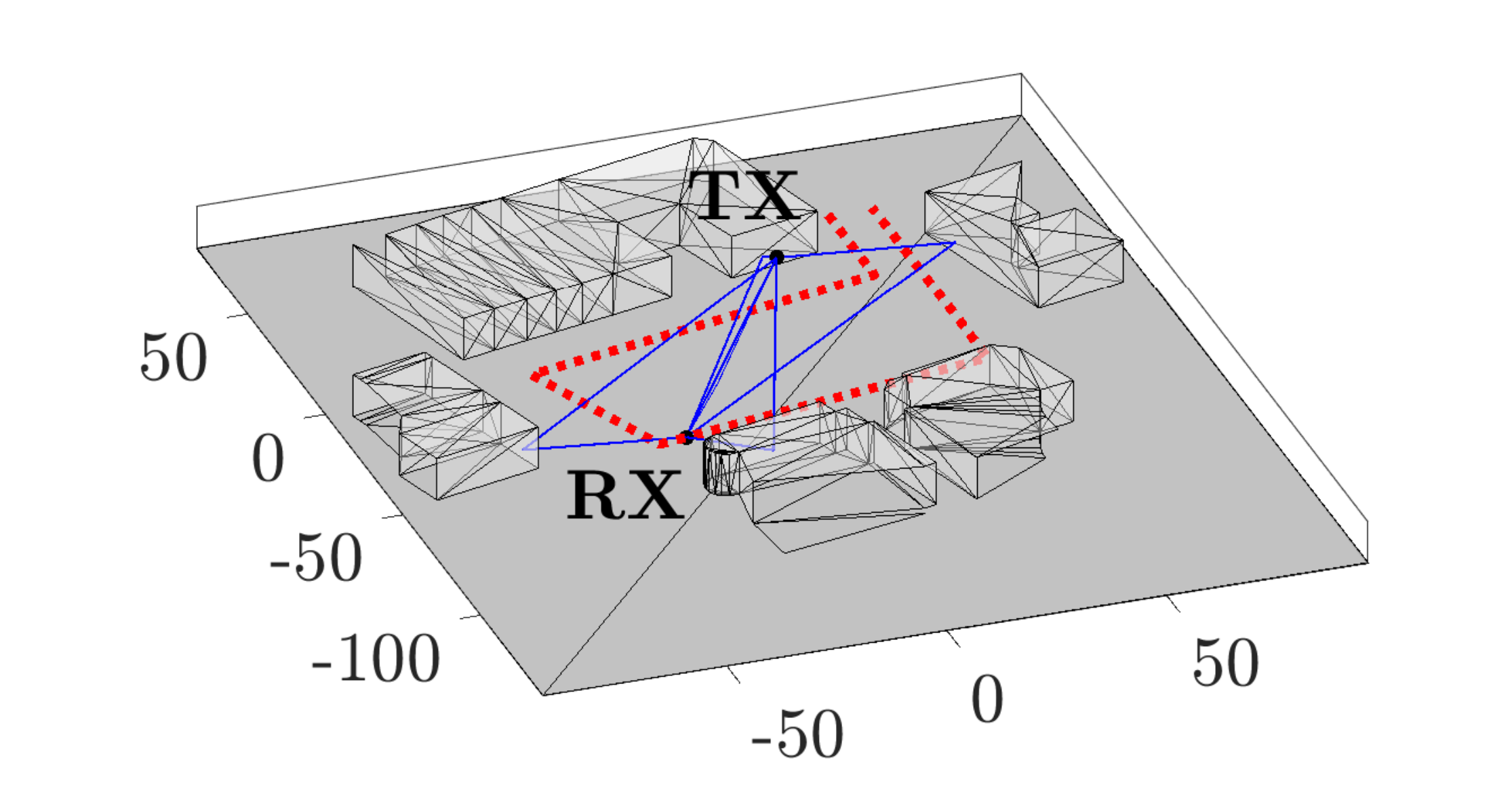}
        \caption{\textit{Parking Lot}}
        \label{fig:parking-lot}
  \end{subfigure}
  \hfill
  \caption{Visual representations of the proposed scenarios.
  Distance measured in meters.}
  \label{fig:scenarios}
  \vspace{-2ex}
\end{figure*}%

\begin{table}[t]
  \caption{Parameters used for the simulations.}
  \label{tab:sim-params}
  \centering
  \begin{tabular}{ll|ll}
    \toprule
    $P_{\rm TX}$ & $30$~dBm & \gls{tx} Array Config. & $8\times 8$\\
    $f_c$ & $60$~GHz & \gls{rx} Array Config. & $4\times 4$\\
    Noise Figure (F) & $5$~dB & Element pattern & Omni-directional\\
    Bandwidth & $400$~MHz & Element spacing & $\lambda /2$\\
    \bottomrule
  \end{tabular}
  \vspace{-3ex}
\end{table}%

Three scenarios are defined as follows:
\begin{enumerate}
  \item \textit{Indoor1}: A simple scenario of a box-like room (\cref{fig:indoor1}) of size $10$~m $\times 19$~m $\times 3$~m. A \gls{tx} is positioned close to the ceiling at $(5, 0.1, 2.9)$~m while the \gls{rx}, $1.5$~m tall, moves inside the room at a speed of $1.2$~m/s in spiral-like motion;
  \item \textit{L-Room}: An L-shaped hallway, as depicted in Fig.~\ref{fig:l-room}. Similarly to the \textit{Indoor1} scenario, the \gls{rx} at the same height as the previous one moves at a speed of $1.2$~m/s;
  \item \textit{Parking-Lot}: An outdoor scenario with buildings around a parking area of about $120$~m $\times 70$~m, as shown in \cref{fig:parking-lot}. A \gls{tx} is positioned on a building $3$~m high and the \gls{rx} is moving at a speed of $4.17$~m/s ($15$~km/h) around the parking lot.
\end{enumerate}
In each scenario, the location of the moving \gls{rx} node was sampled every $5$~ms, for a total number of timesteps of about $9\,000$, $12\,500$, and $15\, 000$, respectively.
A list of parameters used in our simulations is shown in \cref{tab:sim-params}.
Optimal single-stream SVD-based beamforming is computed at each time step.

The following performance metrics are considered:\footnote{In this paper, we focus on lower-layer performance metrics. In turn, investigating the impact of the proposed \gls{rt} simplifications on higher-layer performance metrics represents a very interesting research topic that will be part of our future work.}
\begin{itemize}
  \item The \gls{rt} simulation time $T_{\rm RT}$ [s], i.e., the time taken by the RT software to compute the channel between each pair of nodes at each time-step;
  \item The MATLAB-based Network Simulator time $T_{\rm NS}$  [s], i.e., the time taken by our custom MATLAB simulator to compute the relevant metrics starting from the output of the RT software;
  \item The \gls{nrmse} of the \gls{snr}, an accuracy indicator that compares the \gls{snr} $\Gamma_t$ experienced when the most accurate RT settings (e.g., with $\eta_{\rm max}=4$ and $\gamma_{\rm th}=-\infty$ for the \textit{L-room} scenario) are considered and the \gls{snr} $\hat{\Gamma}$ experienced when different combinations of \gls{rt} simplifications are applied. Formally, if $\sigma_{\Gamma}$ represents the standard deviation of the baseline SNR $\Gamma$, we have
  \begin{equation}
    \text{NRMSE} =  \frac{\text{RMSE}}{\sigma_{\Gamma}}
    = \frac{\sqrt{\mathbb{E}\left[\left(\Gamma-\hat{\Gamma}\right)^2\right]}}{\sigma_{\Gamma}}.
  \end{equation}
\end{itemize}
\begin{figure}[t]
 \centering
 \setlength\fwidth{0.8\columnwidth} 
 \setlength\fheight{0.33\columnwidth}
 \input{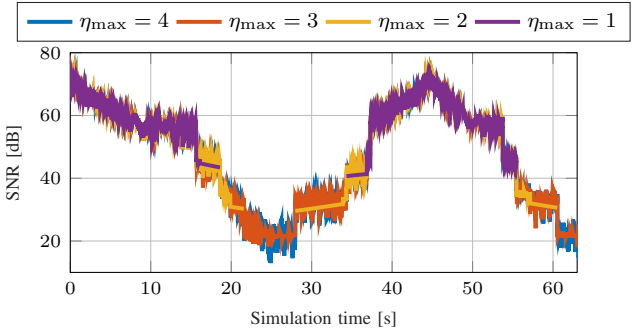}
 \caption{Temporal evolution of the SNR experienced when the test \gls{rx} moves in the \textit{L-room} scenario vs. $\eta_{\rm max}$, fixing $\gamma_{\rm th}=-\infty$.}
 \label{fig:SNR_time_IwcmcLroom}
 \vspace{-2ex}
\end{figure}%
In Fig.~\ref{fig:SNR_time_IwcmcLroom} we consider two nodes, using the \gls{rx} as a probe to monitor the evolution of the \gls{snr} along the trajectory shown in Fig.~\ref{fig:l-room} vs. the maximum number of reflections per \gls{mpc} $\eta_{\rm max}$. 
Rapid variations in the \gls{snr} are due to \glspl{mpc} interfering constructively and destructively, since they travel slightly different path lengths, thus showing a strong fading even for movements in the order of $\lambda=5$~mm (at $60$~GHz). For example, at least 6 of them have similar path gain in the \gls{los} regions, specifically, the \gls{los} ray and the first-order reflections from ceiling, floor, back wall (behind the \gls{tx}), and the two side walls.
Then, we notice that the SNR evolves consistently with the mobility of the \gls{rx}: the SNR suddenly degrades when the \gls{rx} enters a \gls{nlos} condition and is maximized when it is in \gls{los} with its serving \gls{tx}, i.e., around time $t=0$~s and $t=45$~s.
At first glance, it appears that the effect of the \gls{rt} simplifications is not negligible. In particular, considering the lowest possible value of the relative threshold, i.e., $\gamma_{\rm th}=-\infty$, the trend of the SNR visibly changes when progressively limiting the maximum number of reflections for each MPC. The impact of those simplifications is particularly evident when the \gls{rx} operates in \gls{nlos}, i.e., when the number of \glspl{mpc} is as little as one (when fading stops) or even none (when no power is received).
Despite the above considerations, in the following results, we will show more explicitly the accuracy vs. speed trade-off of these parameters in the different scenarios.
Furthermore, we will suggest working points for which computation time is significantly reduced with only minor effects on the accuracy of the model.

\begin{figure}[t]
 \centering
 \setlength\fwidth{0.8\columnwidth} 
 \setlength\fheight{0.33\columnwidth}
 \input{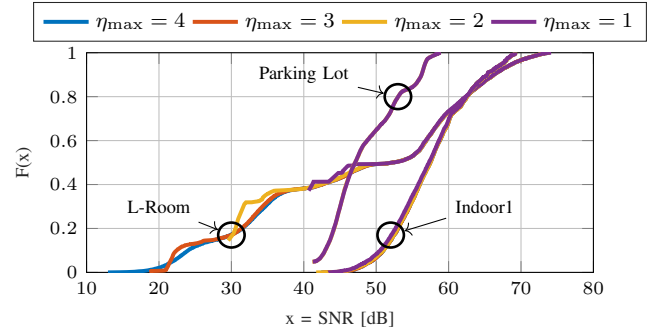}
 \caption{Cumulative Distribution Function of the SNR when the test \gls{rx} moves in different simulation scenarios vs. $\eta_{\rm max}$, with $\gamma_{\rm th}=-\infty$.}
 \label{fig:SNR_cdf_allCampaigns}
 \vspace{-2ex}
\end{figure}

\begin{figure*}[t!]
  \hfill
  \begin{subfigure}[b]{0.45\textwidth}
      \centering
      \setlength\fwidth{0.75\columnwidth} 
      \setlength\fheight{0.33\columnwidth}
%
%
\begin{tikzpicture}

\pgfplotsset{every tick label/.append style={font=\scriptsize}}

\begin{axis}[%
width=\fwidth,
height=0.969\fheight,
at={(0\fwidth,0\fheight)},
scale only axis,
unbounded coords=jump,
clip=false,
xmin=0.6,
xmax=4.6,
xlabel style={font=\scriptsize\color{white!15!black}},
xlabel={Number of Reflections $\eta_{\rm max}$},
xtick={1,2,3,4},
separate axis lines,
ymin=0,
ymax=1000,
ymode=log,
log ticks with fixed point,
ylabel style={font=\scriptsize\color{white!15!black}},
y tick label style={/pgf/number format/1000 sep=\,},
ylabel={RT Sim. Time [min]},
ytick pos=left,
axis background/.style={fill=white},
ymajorgrids,
yminorgrids,
xmajorgrids,
legend style={legend cell align=left, align=left, draw=white!15!black}
]
\addplot [color=black, forget plot]
  table[row sep=crcr]{%
0.9	1.84990505\\
0.9	2.03111725\\
};
\addplot [color=black, forget plot]
  table[row sep=crcr]{%
1.9	5.33630758333333\\
1.9	9.74582695\\
};
\addplot [color=black, forget plot]
  table[row sep=crcr]{%
2.9	41.54186525\\
2.9	56.97883485\\
};
\addplot [color=black, forget plot]
  table[row sep=crcr]{%
3.9	519.687833583333\\
3.9	560.91238805\\
};
\addplot [color=black, line width=4.0pt, forget plot]
  table[row sep=crcr]{%
0.9	1.90075864166667\\
0.9	2.01137175\\
};
\addplot [color=black, line width=4.0pt, forget plot]
  table[row sep=crcr]{%
1.9	6.82107454166667\\
1.9	9.510534275\\
};
\addplot [color=black, line width=4.0pt, forget plot]
  table[row sep=crcr]{%
2.9	43.3822661333333\\
2.9	55.663775675\\
};
\addplot [color=black, line width=4.0pt, forget plot]
  table[row sep=crcr]{%
3.9	524.764681408333\\
3.9	557.4574266\\
};
\addplot [color=black, line width=1.5pt, draw=none, mark=*, mark options={solid, fill=black, black}, forget plot]
  table[row sep=crcr]{%
0.9	1.97161924166667\\
};
\addplot [color=black, line width=1.5pt, draw=none, mark=*, mark options={solid, fill=black, black}, forget plot]
  table[row sep=crcr]{%
1.9	8.79054155\\
};
\addplot [color=black, line width=1.5pt, draw=none, mark=*, mark options={solid, fill=black, black}, forget plot]
  table[row sep=crcr]{%
2.9	49.7856917583333\\
};
\addplot [color=black, line width=1.5pt, draw=none, mark=*, mark options={solid, fill=black, black}, forget plot]
  table[row sep=crcr]{%
3.9	541.921997191667\\
};
\addplot [color=black, line width=1pt, draw=none, mark=*, mark options={solid, fill=white, draw=black}, forget plot]
  table[row sep=crcr]{%
0.9	1.97161924166667\\
};
\addplot [color=black, line width=1pt, draw=none, mark=*, mark options={solid, fill=white, draw=black}, forget plot]
  table[row sep=crcr]{%
1.9	8.79054155\\
};
\addplot [color=black, line width=1pt, draw=none, mark=*, mark options={solid, fill=white, draw=black}, forget plot]
  table[row sep=crcr]{%
2.9	49.7856917583333\\
};
\addplot [color=black, line width=1pt, draw=none, mark=*, mark options={solid, fill=white, draw=black}, forget plot]
  table[row sep=crcr]{%
3.9	541.921997191667\\
};
\addplot [color=black, line width=1.5pt, draw=none, mark size=2.0pt, mark=o, mark options={solid, black}, forget plot]
  table[row sep=crcr]{%
nan	nan\\
};
\addplot [color=black, line width=1.5pt, draw=none, mark size=2.0pt, mark=o, mark options={solid, black}, forget plot]
  table[row sep=crcr]{%
nan	nan\\
};
\addplot [color=black, line width=1.5pt, draw=none, mark size=2.0pt, mark=o, mark options={solid, black}, forget plot]
  table[row sep=crcr]{%
nan	nan\\
};
\addplot [color=black, line width=1.5pt, draw=none, mark size=2.0pt, mark=o, mark options={solid, black}, forget plot]
  table[row sep=crcr]{%
nan	nan\\
};
\node[right, align=left, rotate=90]
at (0.861cm,-0.37cm) {};
\node[right, align=left, rotate=90]
at (3.732cm,-0.37cm) {};
\node[right, align=left, rotate=90]
at (6.603cm,-0.37cm) {};
\node[right, align=left, rotate=90]
at (9.473cm,-0.37cm) {};
\end{axis}

\begin{axis}[%
width=\fwidth,
height=0.969\fheight,
at={(0\fwidth,0\fheight)},
scale only axis,
unbounded coords=jump,
clip=false,
xmin=0.6,
xmax=4.6,
xtick={\empty},
separate axis lines,
ymin=15,
ymax=85,
ylabel style={font=\scriptsize\color{red}},
ylabel={MATLAB Net. Sim. Time [s]},
yticklabel pos=right,
ytick pos=right,
y tick style={color=red},
every y tick label/.append style={red},
legend style={legend cell align=left, align=left, draw=white!15!black}
]
\addplot [color=red, forget plot]
  table[row sep=crcr]{%
1.1	17.680693\\
1.1	18.413954\\
};
\addplot [color=red, forget plot]
  table[row sep=crcr]{%
2.1	18.078035\\
2.1	27.875633\\
};
\addplot [color=red, forget plot]
  table[row sep=crcr]{%
3.1	20.727978\\
3.1	50.687296\\
};
\addplot [color=red, forget plot]
  table[row sep=crcr]{%
4.09999999999999	22.427066\\
4.09999999999999	78.864264\\
};
\addplot [color=red, line width=4.0pt, forget plot]
  table[row sep=crcr]{%
1.1	17.8042795\\
1.1	18.1732565\\
};
\addplot [color=red, line width=4.0pt, forget plot]
  table[row sep=crcr]{%
2.1	21.849794\\
2.1	27.874629\\
};
\addplot [color=red, line width=4.0pt, forget plot]
  table[row sep=crcr]{%
3.1	24.9011365\\
3.1	48.424732\\
};
\addplot [color=red, line width=4.0pt, forget plot]
  table[row sep=crcr]{%
4.09999999999999	26.6945165\\
4.09999999999999	66.9653455\\
};
\addplot [color=black, line width=1.5pt, draw=none, mark=*, mark options={solid, fill=black, red}, forget plot]
  table[row sep=crcr]{%
1.1	17.9302125\\
};
\addplot [color=black, line width=1.5pt, draw=none, mark=*, mark options={solid, fill=black, red}, forget plot]
  table[row sep=crcr]{%
2.1	26.747589\\
};
\addplot [color=black, line width=1.5pt, draw=none, mark=*, mark options={solid, fill=black, red}, forget plot]
  table[row sep=crcr]{%
3.1	37.6182315\\
};
\addplot [color=black, line width=1.5pt, draw=none, mark=*, mark options={solid, fill=black, red}, forget plot]
  table[row sep=crcr]{%
4.1	43.014197\\
};
\addplot [color=black, line width=1pt, draw=none, mark=*, mark options={solid, fill=white, draw=red}, forget plot]
  table[row sep=crcr]{%
1.1	17.9302125\\
};
\addplot [color=black, line width=1pt, draw=none, mark=*, mark options={solid, fill=white, draw=red}, forget plot]
  table[row sep=crcr]{%
2.1	26.747589\\
};
\addplot [color=black, line width=1pt, draw=none, mark=*, mark options={solid, fill=white, draw=red}, forget plot]
  table[row sep=crcr]{%
3.1	37.6182315\\
};
\addplot [color=black, line width=1pt, draw=none, mark=*, mark options={solid, fill=white, draw=red}, forget plot]
  table[row sep=crcr]{%
4.1	43.014197\\
};
\end{axis}
\end{tikzpicture}%
      \caption{Box-plots representing the computation time vs. $\eta_{\rm max}$ when a test \gls{rx} moves in the \textit{L-room} scenario for all values of $\gamma_{\rm th}$.
      Each box includes every combination of $\gamma_{\rm th}$.}
      \label{fig:runtimeBoxplots_refl}
  \end{subfigure}
  \hfill
  \begin{subfigure}[b]{0.45\textwidth}
      \centering
      \setlength\fwidth{0.75\columnwidth} 
      \setlength\fheight{0.33\columnwidth}
%
%
\begin{tikzpicture}

\pgfplotsset{every tick label/.append style={font=\scriptsize}}

\begin{axis}[%
width=\fwidth,
height=0.969\fheight,
at={(0\fwidth,0\fheight)},
scale only axis,
unbounded coords=jump,
clip=false,
xmin=0.6,
xmax=4.6,
xlabel style={font=\scriptsize\color{white!15!black}},
xlabel={Relative Threshold $\gamma_{\rm th}$ [dB]},
xtick={1,2,3,4},
xticklabels={$-\infty$, $-40$, $-25$, $-15$},
separate axis lines,
ymin=0,
ymax=600,
ylabel style={font=\scriptsize\color{white!15!black}},
ylabel={RT Sim. Time [min]},
ytick pos=left,
axis background/.style={fill=white},
ymajorgrids,
xmajorgrids,
legend style={legend cell align=left, align=left, draw=white!15!black}
]
\addplot [color=black, forget plot]
  table[row sep=crcr]{%
0.9	2.03111725\\
0.9	554.00246515\\
};
\addplot [color=black, forget plot]
  table[row sep=crcr]{%
1.9	1.99162625\\
1.9	560.91238805\\
};
\addplot [color=black, forget plot]
  table[row sep=crcr]{%
2.9	1.95161223333333\\
2.9	529.841529233333\\
};
\addplot [color=black, forget plot]
  table[row sep=crcr]{%
3.9	1.84990505\\
3.9	519.687833583333\\
};
\addplot [color=black, line width=4.0pt, forget plot]
  table[row sep=crcr]{%
0.9	5.8884721\\
0.9	305.49065\\
};
\addplot [color=black, line width=4.0pt, forget plot]
  table[row sep=crcr]{%
1.9	5.633433925\\
1.9	307.630552275\\
};
\addplot [color=black, line width=4.0pt, forget plot]
  table[row sep=crcr]{%
2.9	5.12872686666667\\
2.9	287.532098125\\
};
\addplot [color=black, line width=4.0pt, forget plot]
  table[row sep=crcr]{%
3.9	3.59310631666667\\
3.9	280.614849416667\\
};
\addplot [color=black, line width=1.5pt, draw=none, mark=*, mark options={solid, fill=black, black}, forget plot]
  table[row sep=crcr]{%
0.9	33.3623309\\
};
\addplot [color=black, line width=1.5pt, draw=none, mark=*, mark options={solid, fill=black, black}, forget plot]
  table[row sep=crcr]{%
1.9	31.81197905\\
};
\addplot [color=black, line width=1.5pt, draw=none, mark=*, mark options={solid, fill=black, black}, forget plot]
  table[row sep=crcr]{%
2.9	26.7642542583333\\
};
\addplot [color=black, line width=1.5pt, draw=none, mark=*, mark options={solid, fill=black, black}, forget plot]
  table[row sep=crcr]{%
3.9	23.4390864166667\\
};
\addplot [color=black, line width=1pt, draw=none, mark=*, mark options={solid, fill=white, draw=black}, forget plot]
  table[row sep=crcr]{%
0.9	33.3623309\\
};
\addplot [color=black, line width=1pt, draw=none, mark=*, mark options={solid, fill=white, draw=black}, forget plot]
  table[row sep=crcr]{%
1.9	31.81197905\\
};
\addplot [color=black, line width=1pt, draw=none, mark=*, mark options={solid, fill=white, draw=black}, forget plot]
  table[row sep=crcr]{%
2.9	26.7642542583333\\
};
\addplot [color=black, line width=1pt, draw=none, mark=*, mark options={solid, fill=white, draw=black}, forget plot]
  table[row sep=crcr]{%
3.9	23.4390864166667\\
};
\addplot [color=black, line width=1.5pt, draw=none, mark size=2.0pt, mark=o, mark options={solid, black}, forget plot]
  table[row sep=crcr]{%
nan	nan\\
};
\addplot [color=black, line width=1.5pt, draw=none, mark size=2.0pt, mark=o, mark options={solid, black}, forget plot]
  table[row sep=crcr]{%
nan	nan\\
};
\addplot [color=black, line width=1.5pt, draw=none, mark size=2.0pt, mark=o, mark options={solid, black}, forget plot]
  table[row sep=crcr]{%
nan	nan\\
};
\addplot [color=black, line width=1.5pt, draw=none, mark size=2.0pt, mark=o, mark options={solid, black}, forget plot]
  table[row sep=crcr]{%
nan	nan\\
};
\node[right, align=left, rotate=90]
at (0.861cm,-0.714cm) {};
\node[right, align=left, rotate=90]
at (3.732cm,-0.714cm) {};
\node[right, align=left, rotate=90]
at (6.603cm,-0.714cm) {};
\node[right, align=left, rotate=90]
at (9.473cm,-0.714cm) {};
\end{axis}

\begin{axis}[%
width=\fwidth,
height=0.969\fheight,
at={(0\fwidth,0\fheight)},
scale only axis,
unbounded coords=jump,
clip=false,
xmin=0.6,
xmax=4.6,
xtick={\empty},
separate axis lines,
ymin=15,
ymax=85,
ylabel style={font=\scriptsize\color{red}},
ylabel={MATLAB Net. Sim. Time [s]},
yticklabel pos=right,
ytick pos=right,
y tick style={color=red},
every y tick label/.append style={red},
legend style={legend cell align=left, align=left, draw=white!15!black}
]
\addplot [color=red, forget plot]
  table[row sep=crcr]{%
1.09999999999999	18.413954\\
1.09999999999999	78.864264\\
};
\addplot [color=red, forget plot]
  table[row sep=crcr]{%
2.1	17.932559\\
2.1	55.066427\\
};
\addplot [color=red, forget plot]
  table[row sep=crcr]{%
3.1	17.927866\\
3.1	30.961967\\
};
\addplot [color=red, forget plot]
  table[row sep=crcr]{%
4.1	17.680693\\
4.1	22.427066\\
};
\addplot [color=red, line width=4.0pt, forget plot]
  table[row sep=crcr]{%
1.09999999999999	23.1447935\\
1.09999999999999	64.77578\\
};
\addplot [color=red, line width=4.0pt, forget plot]
  table[row sep=crcr]{%
2.1	22.903092\\
2.1	50.6142975\\
};
\addplot [color=red, line width=4.0pt, forget plot]
  table[row sep=crcr]{%
3.1	21.7747095\\
3.1	30.018131\\
};
\addplot [color=red, line width=4.0pt, forget plot]
  table[row sep=crcr]{%
4.1	17.879364\\
4.1	21.577522\\
};
\addplot [color=black, line width=1.5pt, draw=none, mark=*, mark options={solid, fill=black, red}, forget plot]
  table[row sep=crcr]{%
1.1	39.2814645\\
};
\addplot [color=black, line width=1.5pt, draw=none, mark=*, mark options={solid, fill=black, red}, forget plot]
  table[row sep=crcr]{%
2.1	37.0178965\\
};
\addplot [color=black, line width=1.5pt, draw=none, mark=*, mark options={solid, fill=black, red}, forget plot]
  table[row sep=crcr]{%
3.1	27.347924\\
};
\addplot [color=black, line width=1.5pt, draw=none, mark=*, mark options={solid, fill=black, red}, forget plot]
  table[row sep=crcr]{%
4.1	19.4030065\\
};
\addplot [color=black, line width=1pt, draw=none, mark=*, mark options={solid, fill=white, draw=red}, forget plot]
  table[row sep=crcr]{%
1.1	39.2814645\\
};
\addplot [color=black, line width=1pt, draw=none, mark=*, mark options={solid, fill=white, draw=red}, forget plot]
  table[row sep=crcr]{%
2.1	37.0178965\\
};
\addplot [color=black, line width=1pt, draw=none, mark=*, mark options={solid, fill=white, draw=red}, forget plot]
  table[row sep=crcr]{%
3.1	27.347924\\
};
\addplot [color=black, line width=1pt, draw=none, mark=*, mark options={solid, fill=white, draw=red}, forget plot]
  table[row sep=crcr]{%
4.1	19.4030065\\
};
\end{axis}
\end{tikzpicture}%
      \caption{Box-plots of the computation time vs. $\gamma_{\rm th}$ when a test \gls{rx} moves in the \textit{L-room} scenario for all values of $\eta_{\rm max}$.
      Each box includes every combination of $\eta_{\rm max}$.}
      \label{fig:runtimeBoxplots_relTh}
  \end{subfigure}
  \hfill
  \caption{Each box of the box-plot is delimited by the first and the third quartiles of the simulation time, the box's center dot represents the median of the simulation time, and the lines (\textit{whiskers}) extend from the box towards the minimum and the maximum values.}
  \label{fig:runtimeBoxplots}
  \vspace{-2ex}
\end{figure*}
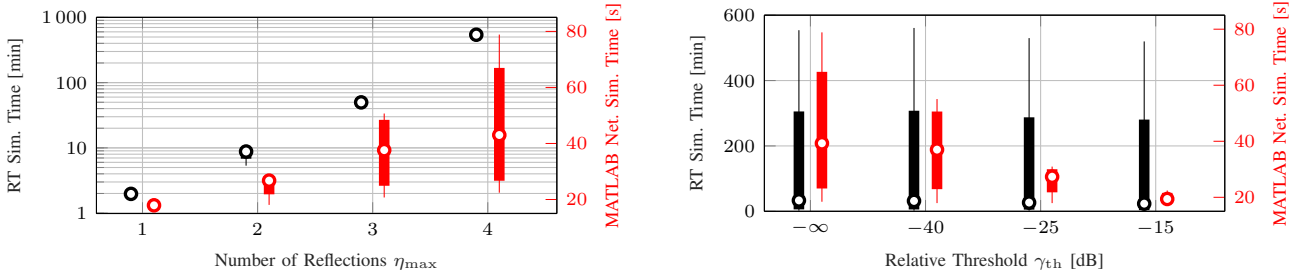

\begin{figure*}[t!]
  \hfill
  \begin{subfigure}[b]{0.45\textwidth}
      \centering
      \setlength\fwidth{0.8\columnwidth} 
      \setlength\fheight{0.33\columnwidth}
%
%
\definecolor{mycolor1}{rgb}{0.85000,0.32500,0.09800}%
\definecolor{mycolor2}{rgb}{0.00000,0.44700,0.74100}%
\definecolor{mycolor3}{rgb}{0.92900,0.69400,0.12500}%
\definecolor{mycolor4}{rgb}{0.49400,0.18400,0.55600}%
\begin{tikzpicture}

\pgfplotsset{every tick label/.append style={font=\scriptsize}}

\begin{axis}[%
width=0.962\fwidth,
height=\fheight,
at={(0\fwidth,0\fheight)},
scale only axis,
xmin=0,
xmax=0.16,
xlabel style={font=\scriptsize\color{white!15!black}},
xlabel={SNR NRMSE},
xticklabel style={/pgf/number format/fixed, /pgf/number format/precision=3},
scaled x ticks=false,
ymin=1,
ymax=1.9,
ylabel style={font=\scriptsize\color{white!15!black}},
ylabel={Speedup},
axis background/.style={fill=white},
xmajorgrids,
ymajorgrids,
legend style={legend cell align=left, align=left, draw=white!15!black, at={(0.5,1.05)}, anchor=south, font=\footnotesize},
legend columns=3
]
\addplot[only marks, mark=*, mark options={}, mark size=2pt, draw=mycolor1, fill=mycolor1, forget plot] table[row sep=crcr]{%
x	y\\
0	1\\
};
\addplot[only marks, mark=square*, mark options={}, mark size=2pt, draw=mycolor1, fill=mycolor1, forget plot] table[row sep=crcr]{%
x	y\\
0.00743728707070553	1.13072438814807\\
};

\coordinate (opt) at (0.014, 1.55);
\coordinate (15db) at (0.159160932903474, 1.7);

\addplot[only marks, mark=triangle*, mark options={rotate=180}, mark size=2pt, draw=mycolor1, fill=mycolor1, forget plot] table[row sep=crcr]{%
x	y\\
0.0143043659427357	1.35109204609987\\
};
\addplot[only marks, mark=diamond*, mark options={}, mark size=2pt, draw=mycolor1, fill=mycolor1, forget plot] table[row sep=crcr]{%
x	y\\
0.159160932903474	1.50661588402883\\
};
\addplot[only marks, mark=*, mark options={}, mark size=2pt, draw=mycolor2, fill=mycolor2, forget plot] table[row sep=crcr]{%
x	y\\
0.0109276449736613	1.52228205667315\\
};
\addplot[only marks, mark=square*, mark options={}, mark size=2pt, draw=mycolor2, fill=mycolor2, forget plot] table[row sep=crcr]{%
x	y\\
0.0125338235450347	1.54336263301392\\
};
\addplot[only marks, mark=triangle*, mark options={rotate=180}, mark size=2pt, draw=mycolor2, fill=mycolor2, forget plot] table[row sep=crcr]{%
x	y\\
0.0145321772237066	1.61180721727736\\
};
\addplot[only marks, mark=diamond*, mark options={}, mark size=2.pt, draw=mycolor2, fill=mycolor2, forget plot] table[row sep=crcr]{%
x	y\\
0.159160932903474	1.88450585347514\\
};

\addplot[only marks, mark=*, mark options={}, mark size=2pt, draw=mycolor2, fill=mycolor2] table[row sep=crcr]{%
x	y\\
-1 -1\\
};
\addlegendentry{$\eta_{\rm max}=1$}

\addplot[only marks, mark=*, mark options={}, mark size=2pt, draw=black, fill=black] table[row sep=crcr]{%
x	y\\
-1 -1\\
};
\addlegendentry{$\gamma_{\rm th}=-\infty$}

\addplot[only marks, mark=triangle*, mark options={rotate=180}, mark size=2pt, draw=black, fill=black] table[row sep=crcr]{%
x y\\
-1 -1\\
};
\addlegendentry{$\gamma_{\rm th}=-25$ dB}

\addplot[only marks, mark=*, mark options={}, mark size=2pt, draw=mycolor1, fill=mycolor1] table[row sep=crcr]{%
x y\\
-1 -1\\
};
\addlegendentry{$\eta_{\rm max}=2$}

\addplot[only marks, mark=square*, mark options={}, mark size=2pt, draw=black, fill=black] table[row sep=crcr]{%
x y\\
-1 -1\\
};
\addlegendentry{$\gamma_{\rm th}=-40$ dB}

\addplot[only marks, mark=diamond*, mark options={}, mark size=2pt, draw=black, fill=black] table[row sep=crcr]{%
x	y\\
-1 -1\\
};
\addlegendentry{$\gamma_{\rm th}=-15$ dB}

\end{axis}

\node[anchor=west,xshift=.3cm,align=left,font=\tiny] at (opt) {Working\\ point\\($\eta_{\rm max} = 1$,\\$\gamma_{\rm th}<-15$ dB)};

\node[anchor=east,xshift=-.5cm, yshift=-.3cm, label={[rotate=-90]\tiny $\gamma_{\rm th}=-15$ dB}] at (15db) {};

\draw (opt) ellipse (0.25cm and 0.35cm);
\draw (15db) ellipse (0.2cm and 0.9cm);

\end{tikzpicture}%
      \caption{\textit{Parking Lot}}
      \label{fig:corrplot_SNR_IwcmcParkingLot}
  \end{subfigure}
  \hfill
  \begin{subfigure}[b]{0.45\textwidth}
      \centering
      \setlength\fwidth{0.8\columnwidth} 
      \setlength\fheight{0.33\columnwidth}
%
%
\definecolor{mycolor1}{rgb}{0.49400,0.18400,0.55600}%
\definecolor{mycolor2}{rgb}{0.92900,0.69400,0.12500}%
\definecolor{mycolor3}{rgb}{0.85000,0.32500,0.09800}%
\definecolor{mycolor4}{rgb}{0.00000,0.44700,0.74100}%
\begin{tikzpicture}

\pgfplotsset{every tick label/.append style={font=\scriptsize}}

\begin{axis}[%
width=0.962\fwidth,
height=\fheight,
at={(0\fwidth,0\fheight)},
scale only axis,
xmin=0,
xmax=0.07,
xlabel style={font=\scriptsize\color{white!15!black}},
xlabel={SNR NRMSE},
xticklabel style={/pgf/number format/fixed, /pgf/number format/precision=3},
scaled x ticks=false,
ymin=1,
ymax=7,
ylabel style={font=\scriptsize\color{white!15!black}},
ylabel={Speedup},
axis background/.style={fill=white},
xmajorgrids,
ymajorgrids,
legend style={legend cell align=left, align=left, draw=white!15!black, at={(0.5,1.05)}, anchor=south, font=\footnotesize},
legend columns=4
]
\addplot[only marks, mark=*, mark options={}, mark size=2pt, draw=mycolor1, fill=mycolor1, forget plot] table[row sep=crcr]{%
x	y\\
0	1\\
};
\addplot[only marks, mark=square*, mark options={}, mark size=2pt, draw=mycolor1, fill=mycolor1, forget plot] table[row sep=crcr]{%
x	y\\
0.00376768852931434	1.26355876000523\\
};
\addplot[only marks, mark=triangle*, mark options={rotate=180}, mark size=2pt, draw=mycolor1, fill=mycolor1, forget plot] table[row sep=crcr]{%
x	y\\
0.0195137854293121	1.78645449333655\\
};

\coordinate (25db) at (0.0195, 3);

\addplot[only marks, mark=diamond*, mark options={}, mark size=2pt, draw=mycolor1, fill=mycolor1, forget plot] table[row sep=crcr]{%
x	y\\
0.0678649454989078	2.09117499706822\\
};

\coordinate (15db) at (0.068, 4.3);

\addplot[only marks, mark=*, mark options={}, mark size=2pt, draw=mycolor2, fill=mycolor2, forget plot] table[row sep=crcr]{%
x	y\\
0.00241133486577957	2.07193948637909\\
};
\addplot[only marks, mark=square*, mark options={}, mark size=2pt, draw=mycolor2, fill=mycolor2, forget plot] table[row sep=crcr]{%
x	y\\
0.00377017565649103	2.26825982882541\\
};
\addplot[only marks, mark=triangle*, mark options={rotate=180}, mark size=2pt, draw=mycolor2, fill=mycolor2, forget plot] table[row sep=crcr]{%
x	y\\
0.0195138826322649	3.52666504764004\\
};
\addplot[only marks, mark=diamond*, mark options={}, mark size=2pt, draw=mycolor2, fill=mycolor2, forget plot] table[row sep=crcr]{%
x	y\\
0.0678649468480462	4.82782285470139\\
};
\addplot[only marks, mark=*, mark options={}, mark size=2pt, draw=mycolor3, fill=mycolor3, forget plot] table[row sep=crcr]{%
x	y\\
0.00440556172859289	3.93896362595061\\
};
\addplot[only marks, mark=square*, mark options={}, mark size=2pt, draw=mycolor3, fill=mycolor3, forget plot] table[row sep=crcr]{%
x	y\\
0.00439412689807798	3.94315377611048\\
};

\coordinate (opt) at (0.00439412689807798, 3.94315377611048);

\addplot[only marks, mark=triangle*, mark options={rotate=180}, mark size=2pt, draw=mycolor3, fill=mycolor3, forget plot] table[row sep=crcr]{%
x	y\\
0.0195199539514098	4.2919152420268\\
};
\addplot[only marks, mark=diamond*, mark options={}, mark size=2pt, draw=mycolor3, fill=mycolor3, forget plot] table[row sep=crcr]{%
x	y\\
0.0678652202520128	6.09322270247571\\
};
\addplot[only marks, mark=*, mark options={}, mark size=2pt, draw=mycolor4, fill=mycolor4, forget plot] table[row sep=crcr]{%
x	y\\
0.0491604664616308	6.04798739140394\\
};
\addplot[only marks, mark=square*, mark options={}, mark size=2pt, draw=mycolor4, fill=mycolor4, forget plot] table[row sep=crcr]{%
x	y\\
0.0491604664616308	6.21006318273185\\
};
\addplot[only marks, mark=triangle*, mark options={rotate=180}, mark size=2pt, draw=mycolor4, fill=mycolor4, forget plot] table[row sep=crcr]{%
x	y\\
0.0491729418952173	6.21250448457701\\
};
\addplot[only marks, mark=diamond*, mark options={}, mark size=2pt, draw=mycolor4, fill=mycolor4, forget plot] table[row sep=crcr]{%
x	y\\
0.067895825920264	6.30094324387644\\
};
\addplot[only marks, mark=*, mark options={}, mark size=2pt, draw=mycolor4, fill=mycolor4] table[row sep=crcr]{%
x	y\\
0 0\\
};
\addlegendentry{$\eta_{\rm max}=1$}

\addplot[only marks, mark=*, mark options={}, mark size=2pt, draw=mycolor2, fill=mycolor2] table[row sep=crcr]{%
x y\\
0 0\\
};
\addlegendentry{$\eta_{\rm max}=3$}

\addplot[only marks, mark=*, mark options={}, mark size=2pt, draw=black, fill=black] table[row sep=crcr]{%
x y\\
0 0\\
};
\addlegendentry{$\gamma_{\rm th}=-\infty$}

\addplot[only marks, mark=triangle*, mark options={rotate=180}, mark size=2pt, draw=black, fill=black] table[row sep=crcr]{%
x y\\
0 0\\
};
\addlegendentry{$\gamma_{\rm th}=-25$ dB}

\addplot[only marks, mark=*, mark options={}, mark size=2pt, draw=mycolor3, fill=mycolor3] table[row sep=crcr]{%
x	y\\
0 0\\
};
\addlegendentry{$\eta_{\rm max}=2$}

\addplot[only marks, mark=*, mark options={}, mark size=2pt, draw=mycolor1, fill=mycolor1] table[row sep=crcr]{%
x y\\
0 0\\
};
\addlegendentry{$\eta_{\rm max}=4$}

\addplot[only marks, mark=square*, mark options={}, mark size=2pt, draw=black, fill=black] table[row sep=crcr]{%
x	y\\
0 0\\
};
\addlegendentry{$\gamma_{\rm th}=-40$ dB}

\addplot[only marks, mark=diamond*, mark options={}, mark size=2pt, draw=black, fill=black] table[row sep=crcr]{%
x	y\\
0 0\\
};
\addlegendentry{$\gamma_{\rm th}=-15$ dB}

\end{axis}

\node[anchor=west,xshift=.1cm, label={[rotate=-90]\tiny $\gamma_{\rm th}=-25$ dB}] at (25db) {};

\node[anchor=east,xshift=-.5cm, label={[rotate=-90]\tiny $\gamma_{\rm th}=-15$ dB}] at (15db) {};

\node[anchor=south,yshift=.1cm,align=center,font=\tiny] at (opt) {Working\\ point};

\draw (25db) ellipse (0.3cm and 1cm);
\draw (15db) ellipse (0.2cm and 1.4cm);
\draw (opt) ellipse (0.2cm and 0.2cm);

\end{tikzpicture}%
      \caption{\textit{L-room}}
      \label{fig:corrplot_SNR_IwcmcLroom}
  \end{subfigure}
  \hfill
  \caption{Speedup vs. SNR NRMSE for different combinations of the RT simplifications when the \gls{rx} moves in the scenario.
  The speedup is relative to the total campaign time $T_{\rm TOT}$ assuming $N_{\rm runs} = 1\,000$ simulations.}
  \label{fig:corrplots}
  \vspace{-2ex}
\end{figure*}
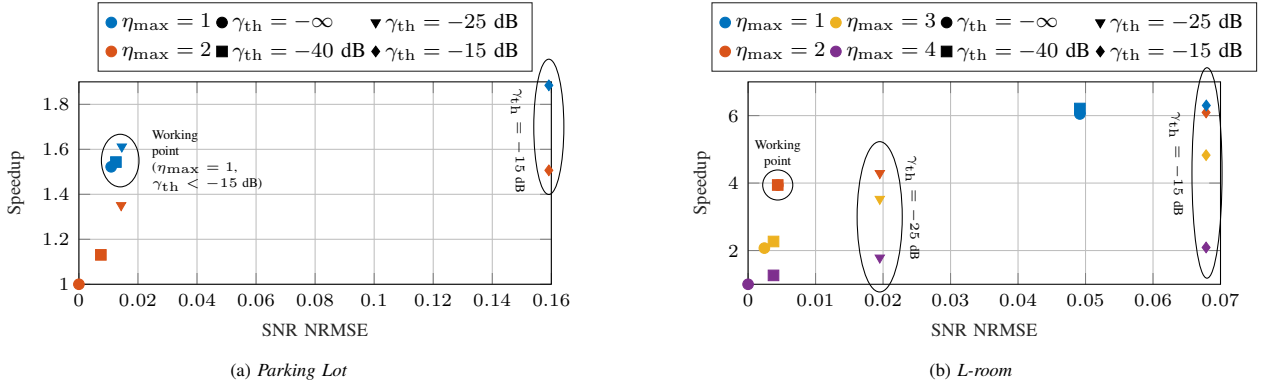

In Fig.~\ref{fig:SNR_cdf_allCampaigns} we plot the \gls{cdf} of the \gls{snr} experienced in the three scenarios as a function of the parameter $\eta_{\rm max}$. 
The abrupt termination of the CDFs for the \textit{L-Room} and \textit{Parking Lot} scenarios is due to positions of the \gls{rx} for which no ray was able to reach it starting from the \gls{tx} position with the given maximum order of reflection $\eta_{\rm max}$, thus resulting in a complete outage.
We observe that, unlike in the \textit{L-room} scenario, in the \textit{Parking Lot} and \textit{Indoor1} scenarios the \gls{rx} preserves the \gls{los} with its serving \gls{tx} for the whole duration of the simulation, thereby maintaining very high values of SNR, i.e., above 40 dB.
Moreover, Fig.~\ref{fig:SNR_cdf_allCampaigns} shows that, while in the \gls{los} regime it is possible to reduce the number of reflections $\eta_{\rm max}$ for each MPC with a minor impact on the accuracy, in the \gls{nlos} regime of the \textit{L-room} scenario (i.e., the leftmost part of the figure) the same operation significantly reshapes the \gls{cdf} of the \gls{snr}, thereby confirming the results we obtained in Fig.~\ref{fig:SNR_time_IwcmcLroom}.

On the other hand, decreasing $\eta_{\rm max}$ may significantly speed up the simulations, as depicted by the box-plots in \cref{fig:runtimeBoxplots}.
We can see that the MATLAB simulation time can be reduced by a factor up to $2.4\times$ going from $\eta_{\rm max}=4$ to $\eta_{\rm max}=1$.
The improvement is even more remarkable considering the \gls{rt} simulation time: the speedup is as significant as $25\times$ considering $\eta_{\rm max}=3$, and even $275\times$ for $\eta_{\rm max}=4$. 
Fig.~\ref{fig:runtimeBoxplots_refl} also shows that the configurations with $\eta_{\rm max}$ set to 3 and 4 exhibit very diverse simulation run time, which is an indication of the increased variability of the channel due to scattering and reflection of the \glspl{mpc} from nearby surfaces.
Similarly, the box-plot in Fig.~\ref{fig:runtimeBoxplots_relTh} illustrates that the speedup factor is proportional to the relative threshold $\gamma_{\rm th}$, since higher values of $\gamma_{\rm th}$ make it possible to reduce the number of \glspl{mpc} to be processed by the ray tracer as described in \cref{sec:simplify}, which represents one of the most computationally intensive steps of both kinds of simulation and which directly impacts on the channel generation process.

Overall, it is possible to identify which level of simplification is most adequate, i.e., provides accurate results while minimizing the overall simulation time.
To this aim, in \cref{fig:corrplots}, we plot the trade-off between simulation \textit{speedup}, defined as the factor by which the simulation time is reduced compared to the baseline, and \textit{NRMSE} of the SNR for the two most complex scenarios, i.e., the outdoor \textit{Parking Lot} and the \textit{L-room}, respectively.
Since, typically, simulations are used to evaluate how changing a set of parameters affects the network performance and should be repeated with several random seeds to increase the robustness of the obtained results, simulation campaigns would reuse the same \gls{rt} channel traces for hundreds or thousands of simulations.
For this reason, we consider the speedup relative to the total campaign time, which is roughly equal to $T_{\rm TOT} = T_{\rm RT} + N_{\rm runs} T_{\rm NS}$, where $T_{\rm RT}$ is the RT computation time, $N_{\rm runs}$ is the number of independent simulations that are run, and $T_{\rm NS}$ is the network simulation run time.

For the \textit{Parking Lot} case (Fig.~\ref{fig:corrplot_SNR_IwcmcParkingLot}), we can see that all the investigated combinations of simplifications with $\gamma_{\rm th}<-15$~dB deliver similar values of SNR NRMSE, while  reducing the computational complexity  with respect to the baseline implementation (i.e., with $\eta_{\rm max}=2$ and $\gamma_{\rm th}=-\infty$).
In this scenario, the measured power has limited contribution from the reflected rays, and the best working point corresponds to the $\eta_{\rm max}=1$ and $\gamma_{\rm th}=-25$~dB configuration: the corresponding speedup is around $60$~$\%$ compared to the baseline ray-tracing model.

For the \textit{L-room} case (Fig.~\ref{fig:corrplot_SNR_IwcmcLroom}), instead, it is possible to identify two operational regimes. On the one hand, very high (low) values of $\gamma_{\rm th}$ ($\eta_{\rm max}$)  would inevitably lead to performance degradation in terms of SNR NRMSE, due to the dominant contribution of the reflected signals to the overall received power.
On the other hand, reflected rays of order higher than the second have a negligible impact in terms of SNR NRMSE.
In this scenario, further reducing $\gamma_{\rm th}$ would result in a considerable increase of the system complexity while leading to negligible accuracy~gain, and the optimal approach would be to select $\eta_{\rm max}=2$  with $\gamma_{\rm th}=-40$~dB instead of the baseline with $\eta_{\rm max}=4$  with $\gamma_{\rm th}=-\infty$~dB, with a speedup equal to 4.

Finally, we highlight that, while limiting the number of \glspl{mpc} reduces the ray tracer's complexity, it may preclude the implementation of beamforming techniques that exploit the sparsity property of the channel to realize simultaneous beams in independent angular direction (e.g., MIMO techniques exploiting spatial multiplexing).
Additionally, while simplifications might have minimal implications on lower-layer performance metrics, e.g., the SNR, their effect on higher-layer metrics, e.g., end-to-end throughput and latency, is still unknown and deserves further investigation.



\section{Conclusions and Future Work}
\label{sec:conclusions}

In this paper, we presented possible simplifications to reduce the complexity of channel modeling through ray tracing. Notably, after introducing the \gls{rt} method based on the \gls{moi}, and the parameters influencing its computational complexity, we discussed two strategies which aim at avoiding computations for \glspl{mpc} which do not contribute significantly to the overall received power. The first limits the maximum reflection order, while the second removes \glspl{mpc} with a path gain which is much smaller than that of the strongest ray. We then evaluated the impact of these simplifications on the \gls{snr}, in three different scenarios, and on the run time of the \gls{rt} and of a custom MATLAB network simulator. We highlighted that, for each scenario, there exists an optimal working point which minimizes the accuracy loss with respect to the baseline, but reduces the channel generation and modeling time by a factor up to four.

In future works, we will consider a more complex \gls{rt}, which also includes diffuse components, according to a quasi-deterministic model~\cite{gentile2018qdDataCenter}. Moreover, we will study the impact of the simplifications on the higher layers of the protocol stack, by using the ns-3 802.11ad module~\cite{assasa11ad2019} (which already integrates the RT) and by extending the ns-3 mmWave module~\cite{mezzavilla2018end} to use RT traces.

\bibliographystyle{IEEEtran}
\bibliography{bibl.bib}

\end{document}